# Mise en place d'une nouvelle habilitation

*D'une solution monolithique vers une architecture en microservice*
D. A. Abboud[a,*] , D. Jacob[a]

*[a] Centre de recherche et d'innovation Talan, Paris*
*[*]Auteur correspondant : david-ahmad.abboud@talan.com (D. A. Abboud)*

## Résumé

*Les applications monolithiques étaient considérées comme la norme en matière de développement de logiciels. Cependant, en raison de l'évolution rapide des technologies et de la demande croissante d'évolutivité et de flexibilité, ces applications sont devenues de plus en plus inadaptées à l'environnement actuel. En réponse à ces défis, les développeurs ont commencé à adopter une architecture en microservice (MS), qui offre une approche modulaire à la création de logiciels. Toutefois, cette transition nécessite une refonte du système d'habilitation afin de répondre aux nouvelles exigences. Deux architectures MS peuvent être déployées : une architecture centralisée ou décentralisée. Face aux exigences des utilisateurs de l'application, c'est une architecture centralisée de la gestion des habilitations qui a été retenue. L'objet de cette étude est d'expliquer le changement d'un système d'habilitation Role-Based Access Control (RBAC) vers une architecture centralisée des habilitations en microservice. La migration est réalisée en 2 étapes : 1) création d'un microservice habilitation 2) Abandon de RBAC*

*Mots clés : Habilitation, Architecture logicielle, Monolithique, Microservice, RBAC.*

## Abstract

*Monolithic applications used to be considered the standard for software development. However, due to the rapid evolution of technology and the increasing demand for scalability and flexibility, these applications have become increasingly inadequate for contemporary environment. In response to these challenges, developers have begun to adopt a microservice (MS) architecture, which offers a modular approach to software creation. However, this transition requires to rethink the enablement system to meet the new requirements. Two MS architectures can be deployed: a centralized or decentralized architecture. Based on the requirements of the application's users, a centralized authorization management architecture was chosen. The purpose of this study is to explain the migration from a Role-Based Access Control (RBAC) authorization system to a centralized microservice authorization architecture. The migration is carried out in two stages: 1) Creation of an authorization microservice 2) Abandonment of RBAC*

## Abréviation

LDAP : Lightweight Directory Access Protocol
MS : Microservice
HS : Habilitation microservice
RBAC : Role-Based Access Control

# 1. Introduction

Les applications modernes ont considérablement augmenté leurs nombres de services rattachés, augmentant de facto leur complexité. Afin de structurer au mieux les flux d'informations qui communiquent entre les services, il est primordial de proposer une architecture robuste et flexible/scalable à la fois, permettant aussi bien un fonctionnement efficient de l'application ainsi qu'une efficience dans ses futures évolutions. Une des architectures historiques consiste à centraliser ces flux d'informations vers une seule et même application. Ce sont les architectures dites monolithiques. Dans un monolithe, toutes modifications de l'application nécessitent de redémarrer l'ensemble de l'application, ce qui rend difficile les phases de développement et de test. Ce type d'architecture limite également les technologies et les langages de programmation utilisés pour développer l'application, qui doivent rester les mêmes pendant toute la durée de vie du projet [1].
Une nouvelle architecture appelée architecture orientée dite " microservice " [2, 3, 4, 5] a été développée pour surmonter ces problèmes. Une architecture microservice (MS) a l'avantage de pouvoir être déployée de manière autonome et de concert avec d'autres applications MS.

Dans le cadre de la refonte d'une application monolithique en une architecture orientée MS, nous avons été amenés à faire évoluer la gestion des habilitations, passant du modèle RBAC (contrôle d'accès basé sur les rôles) à un modèle d'habilitation MS. Ce nouveau modèle de gestion des habilitations doit proposer une solution plus robuste, plus scalable et plus performante.

Plusieurs architectures ont été envisagées pour répondre aux exigences du projet. Ces solutions peuvent être regroupées sous 2 familles :
-       Architectures de type centralisées.
-       Architectures de type décentralisées.

En ce qui concerne les architectures centralisées, nous retrouvons une configuration utilisant un MS dédié aux habilitations (1) et une configuration utilisant le LDAP entreprise à cette fin (2). La solution décentralisée (3) consiste à créer un module dédié aux habilitations pour chaque MS.

En raison des différents avantages et inconvénients évoqués plus haut le projet s'est orienté vers la solution de gestion des habilitations centralisée et la construction d'un MS dédié. Une trajectoire a été pensée de sorte à maintenir l'ancien et le nouveau système en parallèle afin de garantir une transition pérenne pour l'ensemble des utilisateurs.

Pour garantir à notre nouvelle solution un niveau de sécurité équivalent à RBAC, nous devons vérifier que les utilisateurs aient le rôle approprié et que ce dernier ne puisse accéder qu'à ses propres données. Par exemple, si deux utilisateurs d'entreprises différents ont le rôle de " visualisation des comptes", nous devons nous assurer que ce rôle ne donne accès qu'aux données qui appartiennent au bon utilisateur. C'est pourquoi nous devons implémenter une nouvelle fonctionnalité pour vérifier si les données sont accessibles par le bon utilisateur.

## 2. Migration

### 2.1. Ancienne architecture monolithique

#### 2.1.1. RBAC

RBAC pour *Role-Based Access Control* en anglais (contrôle d'accès basé sur les rôles) est un modèle de contrôle d'accès à un système d'information non adapté à la transformation de l'application vers une architecture en MS. Ce modèle fonctionne en définissant les autorisations des utilisateurs en fonction de leur rôle dans l'organisation [6, 7]. Dans un système RBAC, les utilisateurs sont assignés à des rôles, et chaque rôle est associé à un ensemble de permissions qui déterminent ce que les utilisateurs peuvent faire dans le système. Par exemple, un utilisateur peut avoir le rôle de "gestionnaire" et être autorisé à lire et à écrire des données, tandis qu'un autre utilisateur peut avoir le rôle de "lecteur" et ne pas être autorisé à écrire des données [6]. La complexité d'un système RBAC [8] dépend de plusieurs facteurs, notamment de la taille de l'organisation, du nombre de rôles et de permissions, et de la manière dont les utilisateurs et les rôles sont associés. Dans les grandes organisations, le système RBAC peut être très complexe, car il peut y avoir des centaines ou des milliers d'utilisateurs et de rôles différents. De plus, les rôles et les autorisations peuvent changer fréquemment, ce qui peut rendre difficile la gestion des autorisations.

#### 2.1.2. Ancienne architecture monolithique

L'ancien modèle était basé sur la norme RBAC, avec une limitation des habilitations étendue à des entités géographiques (Figure 1). Dans ce modèle, un utilisateur dispose d'habilitations à travers un profil qui est lui-même associé à un nombre de rôles ouvrant des accès à l'application. Chaque association est liée à une entité géographique de sorte à limiter les actions des utilisateurs aux périmètres de cette dernière.

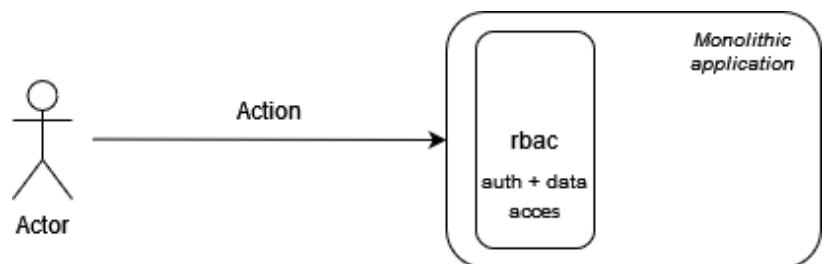

Figure 1 : schéma de l'ancienne architecture monolithique.

Pour des raisons de simplification et afin de faciliter les évolutions à venir, nous avons été amenés à imaginer un nouveau modèle d'habilitation orienté autour des fonctionnalités manipulées par les utilisateurs. Nous avons aussi décidé de nous dissocier des habilitations par entité géographique. Nous avons créé une notion de macro-fonction qui consiste en un regroupement de fonctions. Le nouveau modèle permet toutefois à chaque utilisateur d'associer une maille géographique préférentielle permettant aux applications de filtrer, voire de limiter les accès aux données à cette maille géographique.

## 2.2. Les architectures possibles

Plusieurs architectures ont été envisagées pour répondre aux exigences du projet. Ces solutions peuvent être regroupées en 2 familles :

- Architecture de type centralisé
- Architecture de type décentralisée

L'architecture centralisée utilise un MS dédié pour le processus d'authentification. Cela fonctionne de la manière suivante : chaque MS du système communique avec le MS d'habilitation (Figure 2 - 1) et ceci à chaque fois qu'une nouvelle demande est faite pour vérifier que l'utilisateur ait les rôles d'accès appropriés.

Il y a deux options pour cette architecture : développer un nouveau MS spécifiquement pour ce projet, ce qui permet des solutions personnalisées mais peut être coûteux en temps de développement, ou utiliser une technologie préexistant comme LDAP [9] (Figure 2 - 2). Ce protocole de communication est utilisé pour accéder et gérer les informations stockées dans un annuaire, comme un répertoire d'utilisateurs, de groupes ou d'applications. LDAP est souvent utilisé pour gérer les autorisations dans les systèmes d'information d'entreprise, car il permet d'accéder aux informations d'identification et d'authentification des utilisateurs et de les utiliser pour contrôler l'accès aux ressources. Cependant, utiliser un produit préexistant comme LDAP peut limiter la flexibilité et l'évolutivité du système.

D'un autre côté, l'architecture décentralisée, décrite dans la Figure 2 - 3, attribue des services d'authentification individuels à chaque MS. Cela peut être bénéfique si chaque MS nécessite un processus d'authentification unique, mais peut ne pas être nécessaire si tous les MS ont le même processus et la même structure de données.

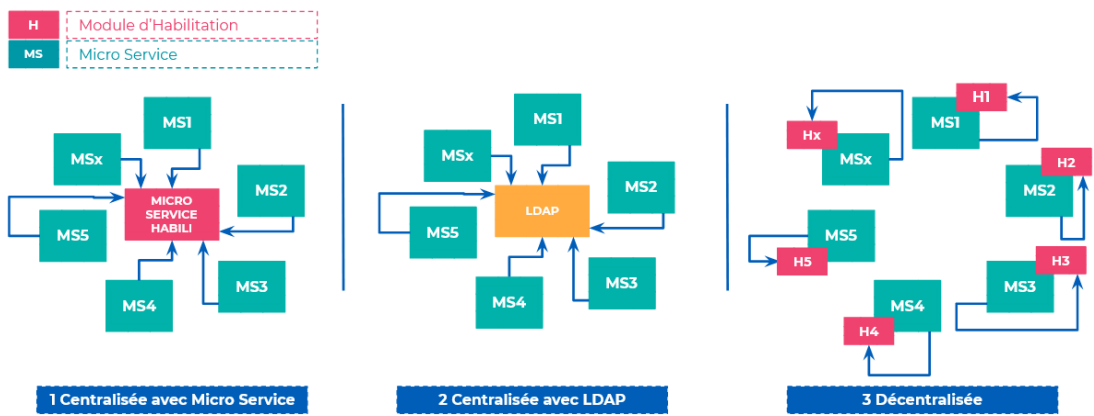

Figure 2 : solutions envisagées avec les architectures centralisées à gauche (1 et 2), décentralisées à droite (3).

## 2.3. Identification de la nouvelle architecture

L'adoption d'une nouvelle architecture doit répondre à plusieurs critères, que nous avons résumé sous la forme de 5 questions :

**1. Est-ce que la nouvelle architecture est facile à mettre en œuvre ?**
Notre solution doit utiliser les nouvelles technologies afin de faciliter l'intégration de cette architecture dans les nouveaux systèmes informatiques.

**2. Est-ce que la nouvelle architecture est facile à modifier ?**

Pour garantir une plus longue durée de vie au produit, il doit être facile à adapter et être capable d'évoluer au fil du temps pour répondre aux nouveaux besoins des utilisateurs et du marché.

**3. Est-ce que cette nouvelle architecture répond au besoin du métier ?**

La nouvelle architecture proposée doit inclure les mêmes fonctionnalités que l'application monolithique afin de garantir une transition facile et transparente pour les utilisateurs.

**4. Quel est le processus d'authentification ?**

Le processus d'authentification doit être clairement définis par les métiers afin de respecter les normes sécurité et protéger les données du système et des utilisateurs.

**5. Est-ce que la solution est personnalisée et flexible ?**

La nouvelle architecture est basée sur les dernières technologies qui facilitent l'intégration, les modifications et la scalabilité du système. On doit être capable de modifier des fonctionnalités facilement sans avoir des régressions qui affectent d'autres MS.

Tableau 1 : récapitulatif des différentes architectures pour le processus d'authentification selon leurs avantages et inconvénients.

| Solution | Centralisée avec microservice | Centralisée avec LDAP | Décentralisée |
|---|---|---|---|
| **Mise en œuvre** | Difficile car cela nécessite de développer un MS spécifique | Moyenne car cela implique l'utilisation d'un produit existant. | Difficile car cela nécessite de développer des services d'authentification individuels pour chaque MS |
| **Modificabilité** | Facile car la solution personnalisée permet une meilleure adaptation aux besoins spécifiques du système. | Difficile car l'utilisation d'un produit préexistant peut limiter la flexibilité et l'évolutivité du système. | Difficile car l'attribution de services d'authentification individuels à chaque MS peut rendre la modification plus complexe. |
| **Réponse au besoin métier** | Oui | Non | Oui |
| **Processus d'authentification** | Un MS dédié pour tous les MS | Un MS dédié pour tous les MS | Services individuels pour chaque MS |
| **Flexibilité** | Meilleure | Limitée | Limitée |

Face aux multiples architectures possibles notre modèle doit pouvoir répondre aux contraintes suivantes :

- Le processus d'authentification doit être simplifié par rapport à la version monolithique.
- Il doit pouvoir lier tous les MS au même modèle d'authentification.
- Utiliser la même structure des données pour tous les processus d'authentification avec les autres MS
- Faciliter l'ajout de nouvelles fonctionnalités et de nouvelles demandes.

En raison des différents avantages et inconvénients évoqués plus haut (Tableau 1) l'architecture retenue est une gestion des habilitations centralisée avec la construction d'un MS dédié. Le développement de cette nouvelle architecture a été mené en parallèle du système actuel afin de garantir une transition pérenne pour l'ensemble des utilisateurs.

2.4. La nouvelle architecture : gestion centralisée avec MS

La transition vers une gestion des habilitations MS s'articule en 2 étapes :
- La première étape consiste à distribuer l'application monolithique sous forme de MS tout en conservant le système RBAC (**Erreur ! Source du renvoi introuvable.**).
- La deuxième étape consiste à se défaire totalement du système RBAC au profit d'une gestion des habilitations MS(**Erreur ! Source du renvoi introuvable.**).

### 2.4.1. Monolithique vers MS

Dans cette phase, l'application monolithique sera remplacée par de multiples MS [5]. Cependant, l'ancien système d'authentification et d'identification est conservé (i.e. le système RBAC). Chaque MS disposera d'un client d'autorisation qui communiquera directement avec le système d'authentification de l'application monolithique par différents canaux en fonction de l'action initiée par l'utilisateur. Une réponse vraie ou fausse est alors envoyée au MS pour déterminer si l'utilisateur a l'accès correct ou non (Figure 3).

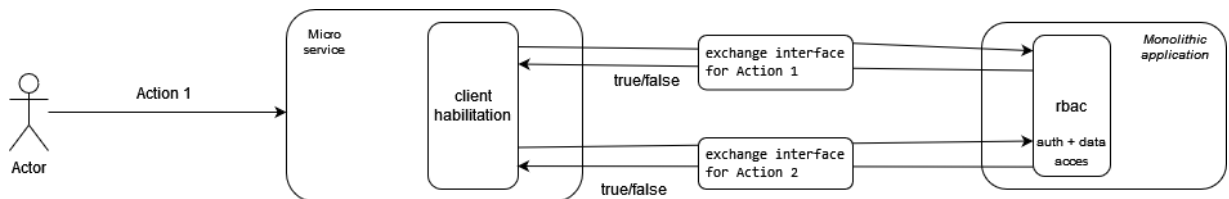

Figure 3 : illustration de la transition d'une architecture monolithique vers des microservices.

### 2.4.2. Suppression du système RBAC

Dans cette seconde phase, une nouvelle application MS dédiée à l'authentification (HS) sera mise en place pour remplacer l'ancien système de l'application monolithique. Chaque MS sera connecté à HS. Si un utilisateur initie une action, le client du MS enverra alors une requête à HS pour récupérer l'ensemble des fonctions macro appartenant à cet utilisateur. Si l'action initiée par l'utilisateur se trouve dans l'ensemble des MS reçus de HS alors l'utilisateur sera authentifié (Figure 4).

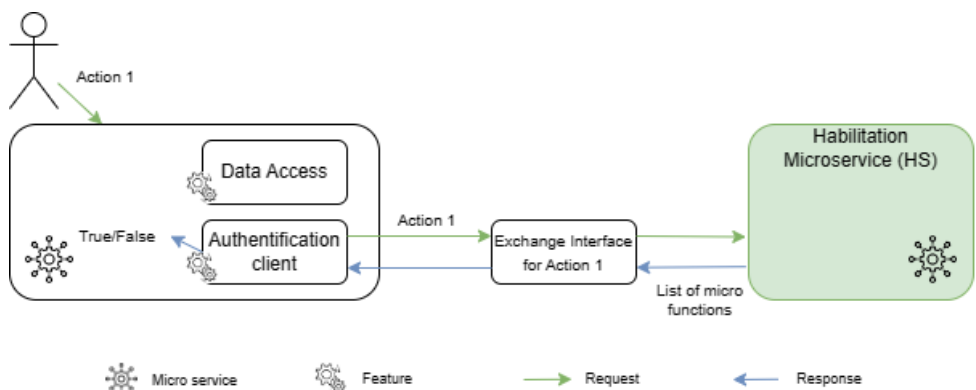

Figure 4 : Illustration de la migration du système vers une habilitation microservice.

Cependant, comme nous l'avons mentionné précédemment, deux utilisateurs de différentes entreprises peuvent avoir la même action, ce qui signifie qu'une nouvelle couche de sécurité doit être mise en place dans chacun de nos MS. Nous appelons cela l'accès aux données ou " *data access* ". Cette partie garantira qu'un utilisateur ayant la bonne action ne peut manipuler que ses propres informations.

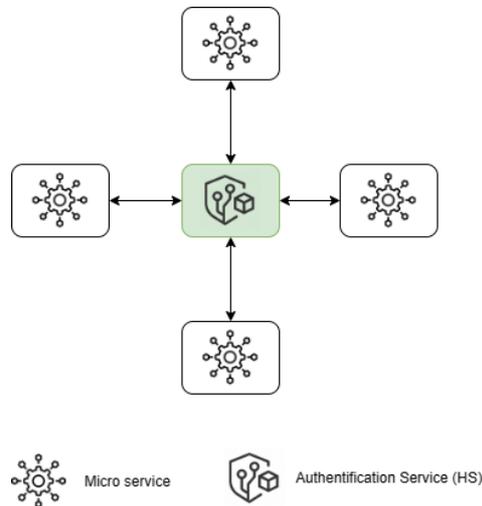

Figure 5 : illustration de l'architecture finale.

La **Erreur ! Source du renvoi introuvable.** illustre la dernière étape de la migration vers un modèle d'habilitation microservice.

Dans la phase finale, l'ensemble de notre application monolithique sera remplacé par des MS. Ces MS communiqueront avec notre nouveau service d'authentification de manière décentralisée.

## 3. Conclusion

L'évolution de notre application monolithique vers un modèle orienté MS a créé un nouveau besoin en matière de gestion des habilitations et des authentifications au sein des différents MS utilisés. La gestion des habilitations pour une application MS est dépendante de l'architecture de ces MS. En effet, deux types d'architectures peuvent être employées : une architecture centralisée ou décentralisée. Ces deux formes diffèrent dans l'utilisation de MS dédiés à l'authentification. L'architecture centralisée utilise un MS dédié pour le processus d'authentification alors que l'architecture décentralisée attribue des services d'authentification individuels à chaque MS.

Le choix s'est porté sur une architecture centralisée car le processus d'authentification doit simplifier le processus d'authentification, lier tous les MS au même modèle d'authentification, utiliser une structure de données commune pour tous les processus d'authentification avec les autres MS, et faciliter l'ajout de nouvelles fonctionnalités et/ou demandes.

La garantie d'un niveau de sécurité équivalent à celui précédemment utilisé avec RBAC est une priorité pour cette nouvelle architecture de MS. Pour atteindre cet objectif, nous avons intégré un système de vérification d'accès aux données dans chaque MS, ce qui permettra de contrôler les autorisations des utilisateurs à différents niveaux du système. Le déploiement de cette nouvelle gestion des habilitations est prévue en 2 étapes : Premièrement distribuer l'application monolithique sous forme de MS tout en conservant le système RBAC et deuxièmement se défaire totalement du système RBAC au profit d'une gestion centralisée des habilitations MS.

Une fois, les phases de développement et de déploiement de notre solution HS terminées, une dernière phase de test et de vérification sera nécessaire pour garantir à la fois, le bon fonctionnement du service, et un niveau de sécurité équivalent à celui utilisé précédemment avec RBAC. Nous devrons également garantir que la mise en place du nouveau système d'authentification n'entraîne pas de détérioration en matière de qualité de service, de temps de vérification et d'authentification.

**Bibliographie**